\documentclass[twocolumn]{aastex63}
\usepackage{url, amssymb, amsmath, graphicx, epsf, fancyhdr, wrapfig, threeparttable}

\newcommand{\xspace}{}

\newcommand{\swift}{{\it Swift}\xspace}

\newcommand{\fermi}{{\it Fermi}\xspace}
\newcommand{\konus}{Konus}
\newcommand{\spiacs}{{\it INTEGRAL} SPI-ACS}
\newcommand{\mcal}{{\it AGILE} MCAL\xspace}

\newcommand{\reffig}[1]{Figure~\ref{fig:#1}}
\newcommand{\reftab}[1]{Table~\ref{tab:#1}}

\newcommand{\p}{\prime}
\newcommand{\mrm}{\mathrm}

\newcommand{\eg}{{e.g.}\xspace}

\newcommand{\Rlow}{$R_{\mathrm{16}}$}
\newcommand{\Rhigh}{$R_{\mathrm{17}}$}

\newcommand{\soft}[1]{\textsc{#1}}

\graphicspath{{./}{figures/}}

\received{}
\revised{}
\accepted{}

\shorttitle{Prompt emission from GRB 221009A}
\shortauthors{Rudolph et al.}

\begin{document}

\title{Multi-messenger model for the prompt emission from GRB 221009A}

\correspondingauthor{Annika Rudolph}
\email{annika.lena.rudolph@nbi.ku.dk}

\author[0000-0003-2040-788X]{Annika Rudolph}
\affil{Niels Bohr International Academy and DARK, 
Niels Bohr Institute, University of Copenhagen, 
Blegdamsvej 17, \\
2100, Copenhagen, Denmark}

\affil{Deutsches Elektronen-Synchrotron DESY, 
Platanenallee 6, 15738 Zeuthen, Germany}

\author[0000-0001-6640-0179]{Maria Petropoulou}
\affil{Department of Physics, 
National and Kapodistrian University of Athens,
University Campus Zografos,
GR 15783, Athens, Greece}

\affil{Institute of Accelerating Systems \& Applications, 
University Campus Zografos, 
GR 15783, Athens, Greece} 

\author[0000-0001-7062-0289]{Walter Winter}
\affil{Deutsches Elektronen-Synchrotron DESY, 
Platanenallee 6, 15738 Zeuthen, Germany}

\author[0000-0001-6536-0320]{\v Zeljka Bo\v snjak}
\affil {Faculty of Electrical Engineering and Computing,
University of Zagreb, 
Unska ul. 3, 
10000 Zagreb, Croatia
}

\begin{abstract}
We present a multi-messenger model for the prompt emission from GRB 221009A within the internal shock scenario. We  consider the time-dependent evolution of the outflow with its impact on the observed light curve from multiple collisions, and the self-consistent generation of the electromagnetic spectrum in synchrotron and inverse Compton-dominated scenarios. Our leptohadronic model includes UHE protons potentially accelerated in the outflow, and their feedback on spectral energy distribution and on the neutrino emission. We find that we can roughly reproduce the observed light curves with an engine with varying ejection velocity of ultra-relativistic material, which has an intermediate quiescent period of about 200 seconds and a variability timescale of $\sim1$~s. We consider baryonic loadings of 3 and 30 that are compatible with the hypothesis that the highest-energetic LHAASO photons might come from UHECR interactions with the extragalactic background light, and the paradigm that energetic GRBs may power the UHECR flux. For these values and the high dissipation radii considered we find consistency with the non-observation of neutrinos and no significant signatures on the electromagnetic spectrum. Inverse Compton-dominated scenarios from the prompt emission are demonstrated to lead to about an order of magnitude higher fluxes in the HE-range; this enhancement is testable by its spectral impact in the Fermi-GBM and LAT ranges. 
\end{abstract}

\keywords{gamma-ray burst: GRB 221009A -- cosmic rays -- neutrinos -- radiation mechanisms: non-thermal}

\section{Introduction} 

Gamma-ray bursts (GRBs) are extremely energetic explosions involving the collapse of a massive star or the merger of two compact objects. As their name suggests, GRBs release most of their electromagnetic output in $\gamma$ rays within a short period of time (ranging from tens of milliseconds to hundreds of seconds). This brief and variable emission, known as the prompt GRB phase, is followed by the afterglow, a long-lasting multi-wavelength emission \citep[for a review see, e.g.][]{2015PhR...561....1K}. The prompt emission is thought to be produced in a relativistic collimated plasma outflow (jet) launched by the central engine via some dissipative mechanism. For kinetically dominated jets, a leading scenario involves energy dissipation and particle acceleration at internal shocks that are produced when portions of the jet are moving outwards with  
varying Lorentz factors \citep[\eg][]{Kobayashi:1997jk, Daigne:1998xc}. 
Part of the remaining jet energy can be later transformed to non-thermal radiation at a relativistic blast wave sweeping up material from the circumburst medium and powering the afterglow \citep{RM1992,CD1999}. 

GRBs are one of the prime targets of multi-messenger astronomy, as they have been detected in gravitational waves \citep[GW~170817/GRB 170817,][]{2017PhRvL.119p1101A} and have been proposed as candidate sources of ultra-high-energy cosmic rays (UHECRs) and astrophysical neutrinos \citep[\eg][]{Waxman:1995vg, Vietri:1995hs}. While the contribution of typical luminosity GRBs during their prompt phase to the diffuse neutrino flux measured by IceCube has been constrained to $\lesssim 1\%$ of the diffuse astrophysical neutrino flux~\citep{IceCube:2022rlk}, the hypothesis that GRBs are UHECR sources cannot be ruled out yet \citep[for a recent review see][]{2022arXiv220206480K}. In fact, GRBs populating the high-end of the isotropic $\gamma$-ray energy distribution ($E_{\gamma,{\rm iso}}>10^{54}$~erg) may provide the necessary energy output per event for powering UHECRs, requiring only a moderate baryonic loading (defined as the energy injected into non-thermal protons versus electrons) and without violating existing neutrino limits \citep{Rudolph:2022ppp}. In addition, energetic bursts may be detected as a single source \citep[\textit{e.g.}][]{Gao:2013fra}.

On October 9th 2022, a very bright GRB was observed at redshift $z = 0.151$ \citep{2022GCN.32648....1D}. The burst triggered the Gamma-Ray Burst Monitor (GBM) on board \fermi \, at 2022-10-09 13:16:59.000~UT~\citep{2022GCN.32636....1V}, about an hour before the detection of a hard X-ray transient by the Burst Alert Telescope (BAT) of the Neil Gehrels \swift \, satellite \citep{GRB221009A_GBM_GCN32642, 2022ATel15650....1D}. The prompt phase of GRB~221009A consisted of a precursor (at about 10~s), followed by an extremely bright emission period about 200~s post GBM trigger. Overall, the prompt emission period lasted roughly 327~s and was composed of several peaks \citep{GRB221009A_GBM_GCN32642}. The preliminary \konus-Wind light curve showed several peaks of roughly 40~s duration \citep{GRB221009A_KONUS_GCN32668}; the initial peaks were separated from the late-time peak by a quiescent period of about 220~s. Some short-timescale variability on the order of seconds might be also visible in the 0.4-100 MeV \mcal \, light curve \citep{GRB221009A_AGILE_GCN32650} and \spiacs \, light curve \footnote{\url{https://grbalpha.konkoly.hu/static/share/GRB221009A_GCN_GRBAlpha.pdf}}. The burst was also observed in high energies (HE, $\gtrsim 0.1$~GeV) by the \fermi \, Large Area Telescope (LAT), starting about 200~s after the GBM trigger (i.e. during the prompt phase) and extending up to $\sim 25$~ks into the afterglow phase~\citep{GRB221009A_LAT_GCN32658}. In addition, very high energy (VHE, $>100$~GeV) photons were  detected by LHAASO, but their association to the prompt phase is not clear given that these photons were observed within a period of up to 2000~s after trigger \citep{GRB221009A_LHAASO_GCN32677}. It has been speculated that the highest energy photons (up to 18~TeV) might come from UHECR interactions with the extragalactic background light (EBL), since such energetic photons escaping the source would be otherwise attenuated by the EBL~\citep{Das:2022gon,AlvesBatista:2022kpg,Mirabal:2022ipw}. These scenarios require a significant amount of energy carried by UHE protons, which might also leave signatures in the electromagnetic spectrum. 

The extreme brightness of this burst caused pile-up in almost all GRB detectors, namely \fermi-GBM and LAT, \textit{KONUS-Wind}, and \textit{AGILE}. For this reason, detailed spectral analysis was unavailable at the time of writing. Preliminary analysis of LAT data (100 MeV - 1 GeV) for $200-800$ s after the GBM trigger provided an estimate of the photon index ($-1.87\pm0.04$) and the photon flux $(6.2 \pm 0.4)\times10^{-3}$~ph cm$^{-2}$ s$^{-1}$ \citep{2022GCN.32658....1P}, but thet time interval excluded from the analysis was recently extended to 300~s \citep{2022GCN.32916....1O}. Hence, the relevance of these results to the main GRB episode detected by GBM is not yet clear.
Moreover, preliminary analysis of \konus \, data during the brightest phase of the event (i.e. $\sim 180-200$~s after the \konus \, trigger) produced a rest-frame peak energy of $E_\mrm{peak} = 1150$~keV and $E_{\gamma,\mathrm{iso}} \simeq 3 \times 10^{54}$~erg.
No associated muon-neutrino track was detected by IceCube in a time range of [-1 hour, +2 hours] from the initial GBM trigger, which resulted in an upper limit on the muon neutrino fluence of $3.9 \times 10^{-2}$ GeV cm$^{-2}$ assuming an $E^{-2}$ neutrino spectrum \citep{GRB221009A_IceCube_GCN32665}. The inferred $\gamma$-ray isotropic energy, the proximity of this event to Earth, and the lack of prompt neutrino detection make GRB~221009A a unique case for the study of multi-messenger signatures from GRBs. 

In this Letter, we present a multi-messenger model for the prompt emission of GRB~221009A. Under the assumption that protons and electrons are accelerated at the fastest possible rate in internal shocks occurring at different radii within the jet, we compute the multi-messenger emission from each collision while taking into account the varying physical conditions in the outflow and the UHECR feedback on both the photon and neutrino emissions. Our goal is to test the hypothesis of UHECR acceleration in the GRB jet by comparing the self-consistently computed broadband photon spectrum and the accompanying neutrino flux with available observational information. Since  details on the prompt spectrum are not yet available, and the mentioned pile-up effects likely introduce some degeneracy, our model also has some predictive power. In this work we indicatively use the following observables: 
\textit{(1)} the peak energy $E_\mrm{peak}=1060$~keV (as fitted for the onset of the bright emission period by \konus), \textit{(2)} the estimated fluence in the GBM band $\mathcal{F}_{\rm GBM} = 2.91 \cdot 10^{-2}$ erg cm$^{-2}$ ($1-1000$~keV) and \textit{(3)} the approximate light curve structure observed by \konus and \spiacs. Peak energy and fluence are reproduced within $\pm 5$ \%.

\begin{table*}[tb]
    \centering
    \caption{Fireball characteristics and microphysics parameters for the different scenarios.}
    \label{tab:input_params}
    \renewcommand{\arraystretch}{1.35}
    \resizebox{1. \textwidth}{!}{%
    \begin{threeparttable}
    \begin{tabular}{lc cc cc }
    \toprule
    &  &  \multicolumn{2}{c}{``\Rlow''-scenario} &  \multicolumn{2}{c}{``\Rhigh''-scenario} \\ 
    Quantity & Symbol  & SYN-dom.  & IC-dom. & SYN-dom.  & IC-dom.   \\ 
        \hline 
    Engine time for main emission period & $t_\mrm{main}$ [s] & \multicolumn{4}{c}{74} \\ 
    Engine quiescent time & $t_\mrm{quiet}$ [s]& \multicolumn{4}{c}{213} \\ 
    Engine time for late-time activity & $t_\mrm{late}$ [s]& \multicolumn{4}{c}{8} \\
    Variability timescale of engine activity & $\delta t_\mrm{var}$ [s]& \multicolumn{4}{c}{1.4} \\
    Averaged $\Gamma$ at beginning & $\langle \Gamma_\mrm{ini} \rangle $ &\multicolumn{2}{c}{ 265 } & \multicolumn{2}{c}{ 663 } \\
    Averaged $\Gamma$ at end & $\langle \Gamma_\mrm{fin} \rangle $&\multicolumn{2}{c}{ 228 }  &  \multicolumn{2}{c}{ 570 } \\
    Averaged $\Gamma$ of emitting plasma & $\langle \Gamma_\mrm{em} \rangle $&\multicolumn{2}{c}{ 293 } &  \multicolumn{2}{c}{ 731 } \\ 
    Averaged radius of emitting plasma & $\langle R_\mrm{Coll} \rangle $ [cm]&\multicolumn{2}{c}{  $1.2 \cdot 10^{16}$} & \multicolumn{2}{c}{ $2.0 \cdot 10^{17}$} \\ 
    Total energy transferred to non-thermal electrons & $E_\mrm{e, NT}$ [erg]  & $4.6 \cdot  10^{54}$   & $9.3 \cdot  10^{54}$ & $4.8 \cdot  10^{54}$& $9.3 \cdot  10^{54}$ \\ 
    Initial fireball kinetic energy & $E_\mrm{kin, ini}$ [erg]  & $5.7 \cdot  10^{56}$   & $9.2 \cdot  10^{56}$ & $3.8 \cdot  10^{57}$& $7.1 \cdot  10^{57}$   \\ 
    Averaged maximal proton energy & $\langle E_\mrm{p, max} \rangle $ [$10^{11}$ GeV]  & 21.9 & 2.7 & 23.7 & 1.1 \\
    Emitted gamma-ray energy (1 - $10^{4}$ keV) & $E_\mrm{\gamma, iso}$ [erg]  & $2.9 \cdot  10^{54}$   & $3.1 \cdot  10^{54}$ & $2.9 \cdot  10^{54}$ & $2.9 \cdot  10^{54}$   \\ 
    \hline 
    Relative fraction of energy transferred to magnetic field & $f_\mrm{B/e}= \epsilon_\mrm{B}/\epsilon_\mrm{e}$ & $1$& $10^{-3}$ & $1$& $10^{-3}$ \\
 Relative fraction of energy transferred to acc. protons & $f_\mrm{p/e}= \epsilon_\mrm{p}/\epsilon_\mrm{e}$ &  3 & 3 & 30 & 30 \\ 
 Power-law index of accelerated electrons & $p_\mrm{e}$ & \multicolumn{4}{c}{2.2} \\
 Minimum Lorentz factor of accelerated electrons & $\gamma_\mrm{e, min}^\prime$ [$10^{4}$]& 3 & 9 & 6 & 11 \\
Power-law index of accelerated protons & $p_\mrm{p}$ & \multicolumn{4}{c}{2.0} \\
Minimum Lorentz factor of accelerated electrons & $\gamma_\mrm{p, min}^\prime$ & \multicolumn{4}{c}{10}  \\
\hline
    \end{tabular}
    \begin{tablenotes}
\item \textit{Notes. --} All listed energies refer to the isotropic equivalent values. Energies and times are reported in the source (engine) frame. Averaged quantities are computed weighing with the shell mass ($\langle \Gamma_\mrm{ini} \rangle $ and $\langle \Gamma_\mrm{fin} \rangle $) or with the dissipated energy of a collision ($\langle \Gamma_\mrm{em} \rangle $, $\langle R_\mrm{Coll} \rangle $, $\langle E_\mrm{p, max} \rangle $). The maximal proton energies are computed considering synchrotron and adiabatic losses and equally refer to the engine frame.
\end{tablenotes}
\end{threeparttable}
}
\end{table*}

\section{Model and implementation}

Our model is fully described in an accompanying paper \citep{Rudolph:2022ppp} to which we refer the interested readers for details. Here we summarize the key points of the model and present its main parameters in Tab.~\ref{tab:input_params}. In short, a relativistic outflow is discretized as shells ejected with different Lorentz factors from a central engine operating over a time $t_{\mathrm{eng}}$ (as measured in the engine rest frame). Shells catch up with each other and merge at a radius $R_{\mathrm{Coll}}$ from the central emitter, where energy is dissipated and particles are accelerated. Shells continue to propagate, merge and dissipate energy until their velocity distribution is such that no more collision occur. The dissipated energy in each inelastic collision, which can be obtained from energy and momentum conservation, is distributed into non-thermal electrons, protons and magnetic fields (for simplicity we assume that no energy remains in thermal plasma). We introduce the partition parameters $f_{\rm p/e}$ (baryonic loading) and $f_{\rm B/e}$ (magnetic loading) describing the ratio between proton/electron and magnetic/electron energy density, respectively (here only referring to accelerated, non-thermal particles). 
Assuming that the sub-MeV prompt spectrum is predominately produced by synchrotron radiation of accelerated electrons, we normalize the fireball kinetic energy to the total energy transferred to non-thermal electrons $E_\mathrm{e, NT}^\mathrm{tot}$ that is needed to produce a given $E_\mrm{\gamma, iso}$. Based on our simulations the energy dissipation efficiency\footnote{The fireball radiative efficiency is then found by multiplying $\varepsilon_{\rm diss}$ with the fraction of energy carried by primary electrons and the synchrotron radiative efficiency.} of the fireball is $\varepsilon_{\mathrm{diss}} \simeq 0.04$, for the specific Lorentz factor distribution assumed here. The required fireball kinetic energy (isotropic equivalent) is then indirectly related to the energy partition parameters and, in  particular, increases with increasing baryonic loading as  
$ E_\mrm{kin,ini} =  \varepsilon_{\rm diss}^{-1} E_\mathrm{e, NT}^\mathrm{tot}(1 + f_\mrm{p/e} + f_\mrm{B/e})$. In contrast to \citet{Rudolph:2022ppp}, we normalize all models to a similar $E_\mrm{\gamma, iso}$, leading to higher $ E_\mrm{kin,ini}$ for cases with a low electron synchrotron efficiency.

\begin{figure*}[t]
    \centering
\includegraphics[width = 0.42 \textwidth]{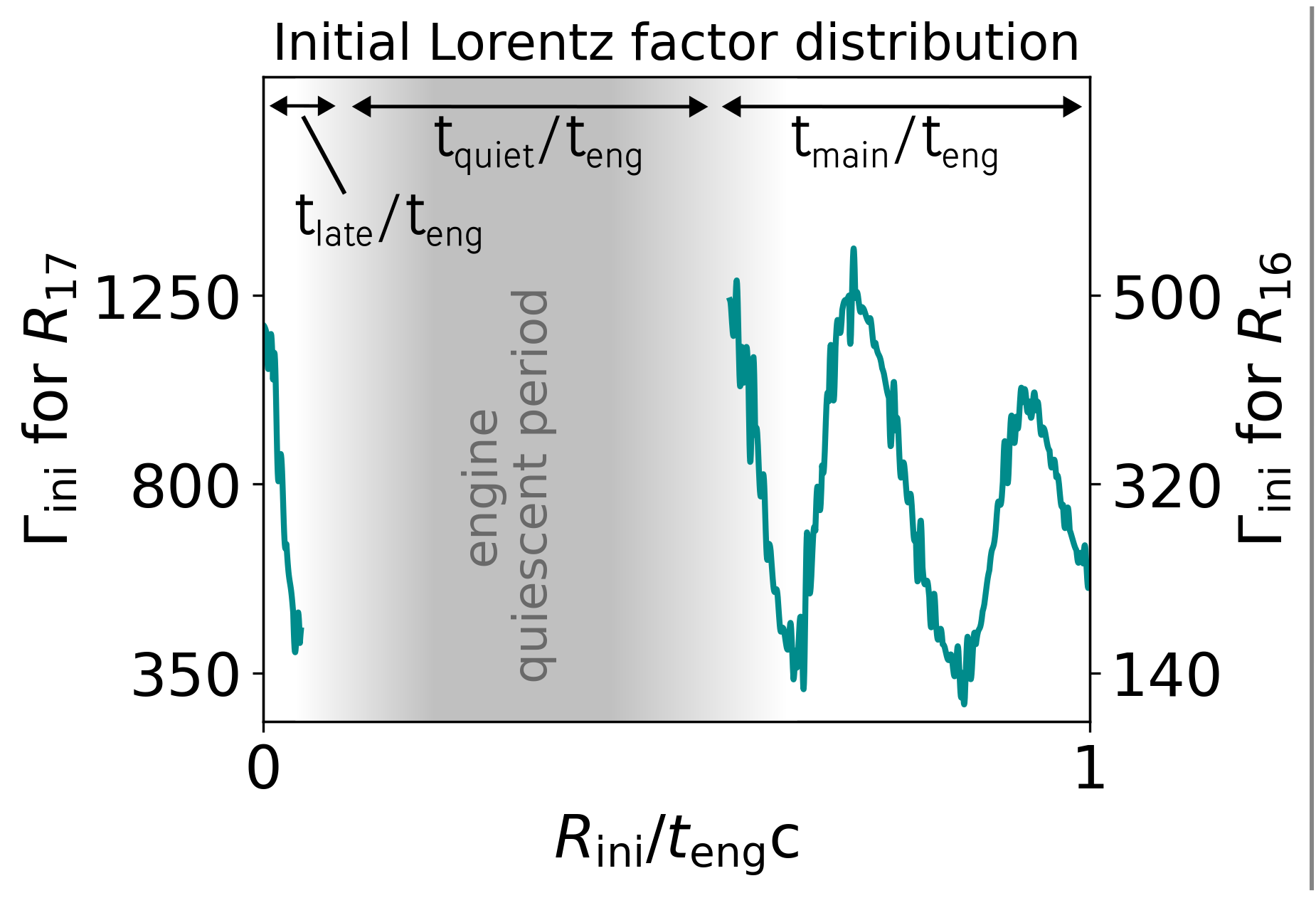}
\includegraphics[width = 0.57\textwidth]{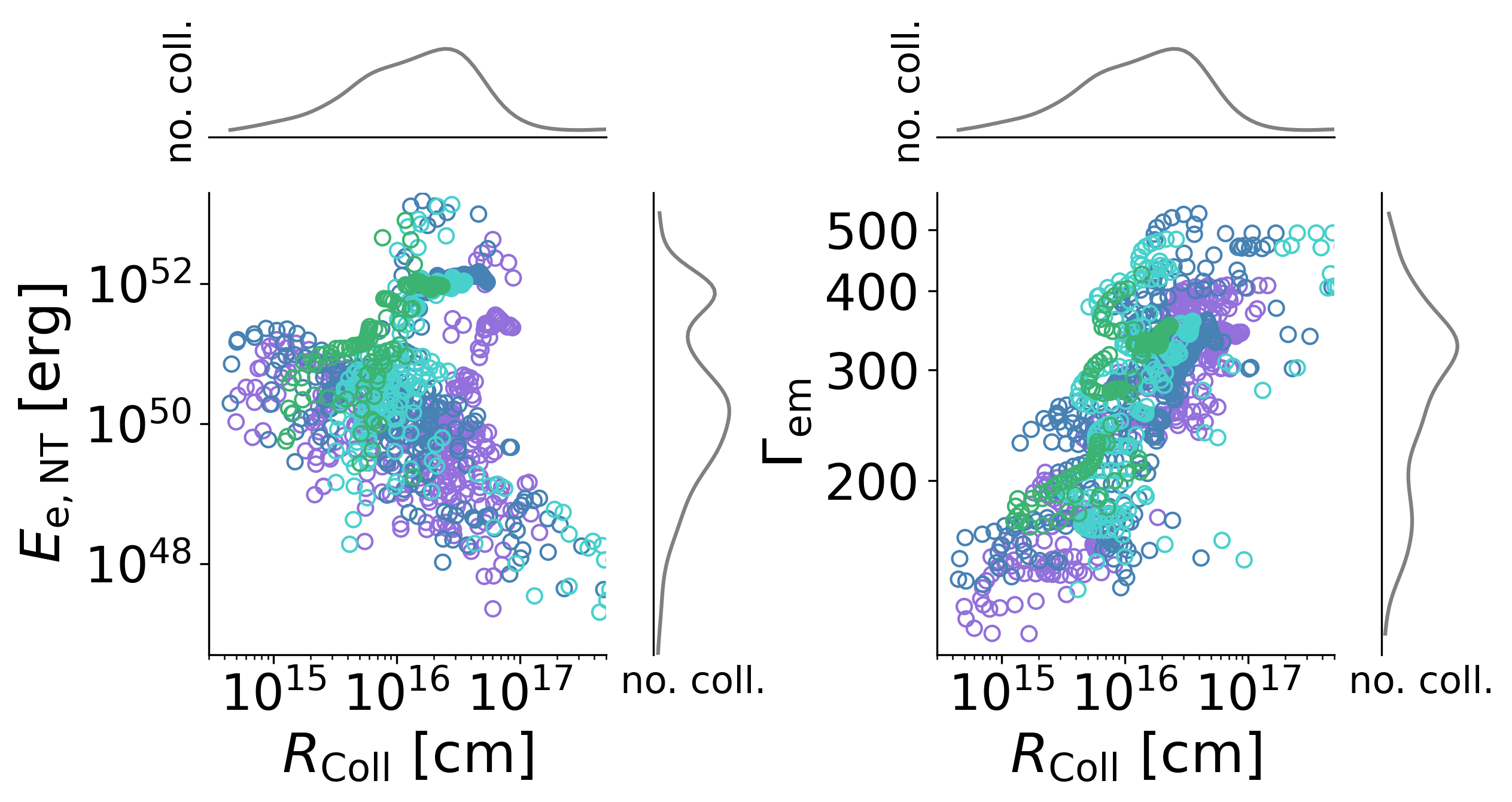}
    \caption{\textit{Left:} Distribution of Lorentz factors of plasma shells launched by the central engine, here $R_\mrm{ini}$ is the initial radius of a plasma shell. Two cases are shown leading to high (left axis) and low (right axis) collision radii. The parameter $t_\mrm{eng}$ represents the total activity time of the engine, i.e., $t_\mrm{eng} = t_\mrm{main} + t_\mrm{quiet} + t_\mrm{late}$.
    \textit{Middle and right:} Two-dimensional phase space of 
    the collision radius and the total energy carried by non-thermal electrons (middle) or the Lorentz factor of the emitting plasma (right). Results are shown for the case of 
    low dissipation radii and SYN-dominated electron cooling (Table~\ref{tab:input_params}). The collisions (each represented by a circle) are separated in color by the time interval during which they will be observed: purple (0--36.5~s), blue (36.5--71.5~s), aquamarine (71.5--129.5~s) and green (from 308.5~s). 
    Grouped collisions correspond to four distinctive pulses of the light curve shown in Fig.~\ref{fig:light_curve}.
    }
    \label{fig:initial_gammas_dist_collisions}
\end{figure*}

To reproduce the structure of the observed light curve we use a varying Lorentz factor profile, see Fig.~\ref{fig:initial_gammas_dist_collisions} (left panel), here without addressing the question which engine properties would lead to such a profile. 
The initial Lorentz factor distribution is obtained in two steps: First, we reproduce the broad structure of the observed light curve (a dip and two bright peaks, followed by a quiescent period and late-time emission) through sine waves with relative amplitudes matching the observed flux variations and durations of these engine activity intervals. The engine activity is correspondingly characterised by three time intervals: The activity time of the main emission period $t_\mrm{main}$, the engine quiescent time $t_\mrm{quiet}$ and activity time for the late-time emission $t_\mrm{late}$.
Second, after inferring a short-time variability timescale of $\delta t_{\mathrm{var}}\simeq 1$s from observations we add modulations on this timescale, with amplitude drawn randomly from a normal distribution with standard deviation $0.08 \cdot \langle \Gamma \rangle$. From 20 realisations of this random process, we select the initial configuration that best matches the observed light curve by eye.

We present two different scenarios for the initial Lorentz factor distribution -- denoted as ``\Rlow'' and ``\Rhigh'' -- that produce collisions at an average radius $\langle R_\mrm{Coll} \rangle \sim 10^{16}$~cm and $\langle R_\mrm{Coll} \rangle \sim 2 \cdot 10^{17}$~cm respectively, and probe higher and lower typical plasma densities. We verified that in both scenarios the bulk of energy dissipation occurs below the estimated deceleration radius for typical parameters of the circumburst medium.
These scenarios are motivated by estimates for the optical thickness of the emitting plasma to $\gamma \gamma$ pair production~\citep{Murase:2022vqf} and the requirement that the bulk of energy is dissipated below the approximate deceleration radius. In principle,
high $\Gamma$ factors are also expected from the empirical $E_{\gamma,\mathrm{iso}}-\Gamma$ relationships~\citep{Ghirlanda:2017opl}. As an example, we show for ``\Rlow'' the two-dimensional distributions of the non-thermal electron energy and the Lorentz factor of the emitting shell with collision radius in the middle and right panels of Fig.~\ref{fig:initial_gammas_dist_collisions}. 

We self-consistently evaluate the emission from non-thermal electrons and protons, accelerated in shocks from the collision of shells, and injected with power-law distributions into the radiation zone, which is the hot plasma of the merged shell. The full-burst emission is then integrated over all collisions, taking into account the curvature of the emitting surface. The maximal electron and proton energies are limited by the dominating energy loss processes assuming efficient acceleration (operating at the Bohm limit). For each scenario, the minimum electron Lorentz factor $\gamma_\mrm{e, min}^\prime$ is set such the peak energy of 1060~keV is reproduced; The minimal proton energies are generally fixed to $\gamma'_\mathrm{p, min} =10$. 
We assume a power-law slope of primary protons of $p_\mrm{p} = 2.0$ as in \citet{Rudolph:2022ppp}. In our model the primary electron power-law slope would usually be determined by the high-energy photon index. As there was no reliable measurement for the high-energy slope at the time of writing, we adapt $p_\mrm{e} = 2.2 $ as suggested for mildly relativistic shocks \citep{Crumley:2018kvf}.
We then numerically solve\footnote{The simulations are performed with the proprietary code {\tt AM$^3$}~\citep{Gao:2016uld}.} the coupled system of integrodifferential equations describing the temporal evolution of particle distributions (electrons/positrons,  photons, protons, neutrons, pions, muons, neutrinos) including the relevant processes, such as synchrotron emission and absorption, inverse Compton scattering (ICS), photo-pair and photo-pion production, $\gamma \gamma$ pair production, adiabatic cooling, and escape. Our approach fully captures the electromagnetic cascade induced by photo-hadronic and $\gamma \gamma$ pair production in each merged shell\footnote{We do not consider interactions of particle populations between shells.}. 

As discussed in detail in \citet{Rudolph:2022ppp}, the electromagnetic spectrum will be, even in the leptohadronic case, dominated by the primary leptonic emission:
a synchrotron component peaking in the MeV band and an inverse Compton component emerging at higher energies (GeV band) for low enough $f_{\rm B/e}$ values. We therefore discuss a scenarios synchrotron (``SYN'')-dominated and inverse Compton (``IC'')-dominated one with different values of $f_{\rm B/e}$ (see  Tab.~\ref{tab:input_params}), characterized by different photon spectra and light curves. 
We also include the effects of EBL attenuation using the model of \citet{Dominguez:2010bv}, calculated with the open-source \soft{gammapy}-package \citep{Deil:2017yey, Nigro:2019hqf}. 

A crucial parameter in every multi-messenger model is the baryonic loading. 
\citet{Rudolph:2022ppp} showed that $f_{\rm p/e} \gtrsim 3$ is required for $E_{\gamma,\mathrm{iso}} \simeq 3 \cdot 10^{54} \, \mathrm{erg}$, if such energetic GRBs ought to power the UHECRs. Much higher baryonic loadings in combination with low $R_{\mathrm{Coll}}$ lead to spectral distortions of the photon spectrum (even in the GBM and LAT bands), and efficient neutrino production, which might be in tension with multi-messenger limits \citep[for details, see Sec.~6 in][]{Rudolph:2022ppp}. A different argument comes from the observed VHE photons in LHAASO, if these were produced  by interactions of UHECRs with the EBL 
\citep{Das:2022gon,AlvesBatista:2022kpg,Mirabal:2022ipw}. \citet{Das:2022gon} derive an isotropic-equivalent $E_{p,\mathrm{iso}} \simeq 3.9 \cdot 10^{54} \, \mathrm{erg}$ for escaping cosmic rays in the range 0.1-100 EeV, which yields $E_{p,\mathrm{iso}} \gtrsim 2 \cdot 10^{55} \, \mathrm{erg}$ bolometrically corrected for our considered energy range, if all UHECRs free-streamingly escape. In our model, this energy budget is met for $f_{\rm p/e} \gtrsim 3$. For \citet{AlvesBatista:2022kpg}, the given fraction into UHECRs is lower, but the bolometric correction factor is much higher, as only UHECRs at the highest energies were considered in that work. In what follows, we choose $f_{\rm p/e} = 3$ for ``\Rlow'' and $f_{\rm p/e} = 30$ for ``\Rhigh''. The former is compatible with the above estimates, while the latter is a more aggressive version  (still compatible with neutrino bounds) that will challenge the GRB energetics in terms of the kinetic energy required.
Purely leptonic results will not be shown, as we found no substantial modifications due to hadronic contributions in the observable photon spectra for the baryonic loadings considered.

\section{Results} 

\begin{figure*}[t]
    \centering
\includegraphics[width = \textwidth]{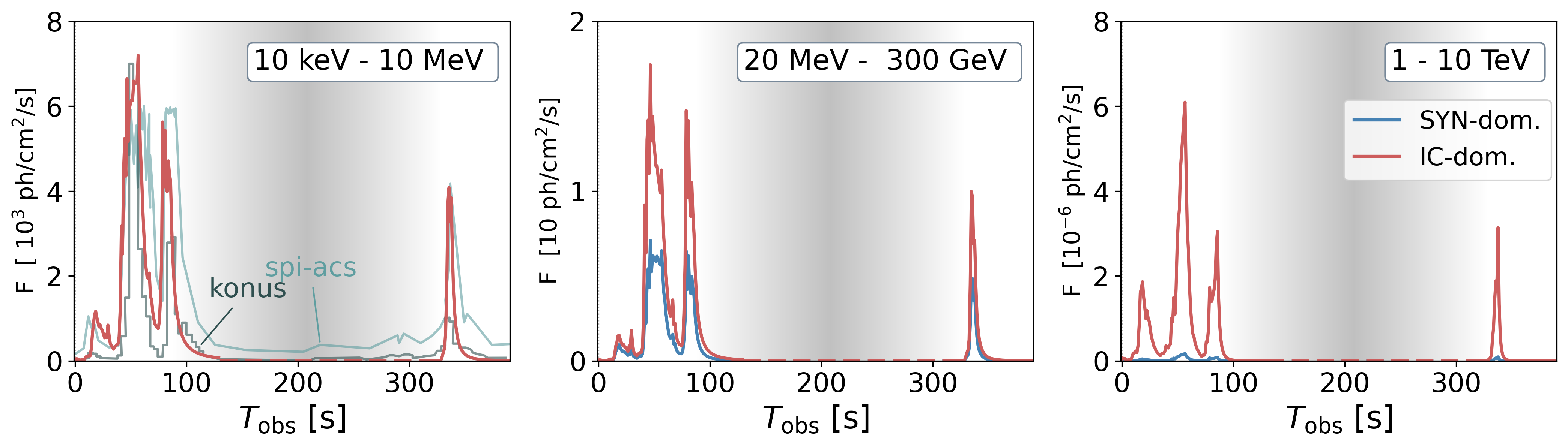}
 \caption{Synthetic light curves for three energy ranges obtained in the ``\Rlow'' scenario for the IC- and SYN-dominated cases (in all panels indicated as red and blue curves respectively, see legend in right panel). In the left panel we show only the IC-dominated case as it resembles the SYN-dominated one. We also add observed light curves of \konus \, and \spiacs \, that operate in a comparable energy range, re-normalised to match the same scale and shifted in time to match the same peak times (-175~s for \konus, and -225~s for \spiacs). We set $T_\mrm{obs} = 0 $ for the synthetic light curve as the minimum collision time in the observer's frame.}
    \label{fig:light_curve}
\end{figure*}

\textit{Light curves.} By construction of our model, the rough structure of  the light curves shown in Fig.~\ref{fig:light_curve} reflects the initial Lorentz factor distribution in Fig.~\ref{fig:initial_gammas_dist_collisions}. Each pulse of the synthetic light curves is formed by shells colliding over a wide range of radii (as can be inferred from the colour coding in  Fig.~\ref{fig:initial_gammas_dist_collisions}). 
Collisions within the same pulse may thus have different opacities, potentially introducing time-lags between different energy ranges.

In the left panel, we indicate the observed light curves of \spiacs \, and \konus, shifted to match the beginning of our synthetic light curve that does not include the precursor emission. 
The general structure of the observed light curve is reproduced by our model, but our results should also be representative for light curves with a similar general structure. The relative intensity of the two bright pulses in the observed light curve may strongly be impacted by pile-up effects in the detector. Here, we assumed that the first bright peak is intrinsically much more energetic than the second one. In our model the relative height of the peak is controlled by the ratio of Lorentz factors of the colliding shells (e.g. a bright peak is produced when this ratio is large). 
Available $\gamma$-ray light curves at the time of writing indicate short-time variability on the second timescale, which was modeled by stochastic variations of $\Gamma_{\rm ini}$ on a timescale $\delta t_\mrm{var} = 1.4$~s (see Fig.~\ref{fig:initial_gammas_dist_collisions}). However, if refined analysis of the light curves revealed variability on a shorter timescale, this would shift the typical collision radius inwards where the densities are higher.
As shown in \citet{Rudolph:2022ppp}, this would lead to stronger signatures of secondary particles, such as secondary leptons and neutrinos, for the same baryonic loading. 

The synthetic HE light curve (shown in the middle panel of \reffig{light_curve}) is very similar to the keV-MeV light curve. Preliminary analysis of LAT data using \soft{gtburst}\footnote{\url{https://fermi.gsfc.nasa.gov/ssc/data/analysis/scitools/gtburst.html}} showed no evidence for a fourth peak in the $20~\rm MeV-300~\rm  GeV$ light curve. If this is later verified by detailed LAT analysis, then late-time collisions with low Lorentz factors will be needed in our model in order to suppress the late-time HE emission due to high internal opacity to $\gamma \gamma$ pair production. While the shape of the HE light curve is similar in the SYN- and IC-dominated scenarios, the HE flux is higher in the latter case, suggesting that primary electrons are cooling more efficiently via ICS (see also next paragraph). Non-negligible VHE emission between 1-10~TeV is also expected in the IC-dominated scenario (right-hand panel), with similar light curve as the in the lower energy bands. 

\begin{figure*}[t]
    \centering
  \includegraphics[width = 0.48 \textwidth]{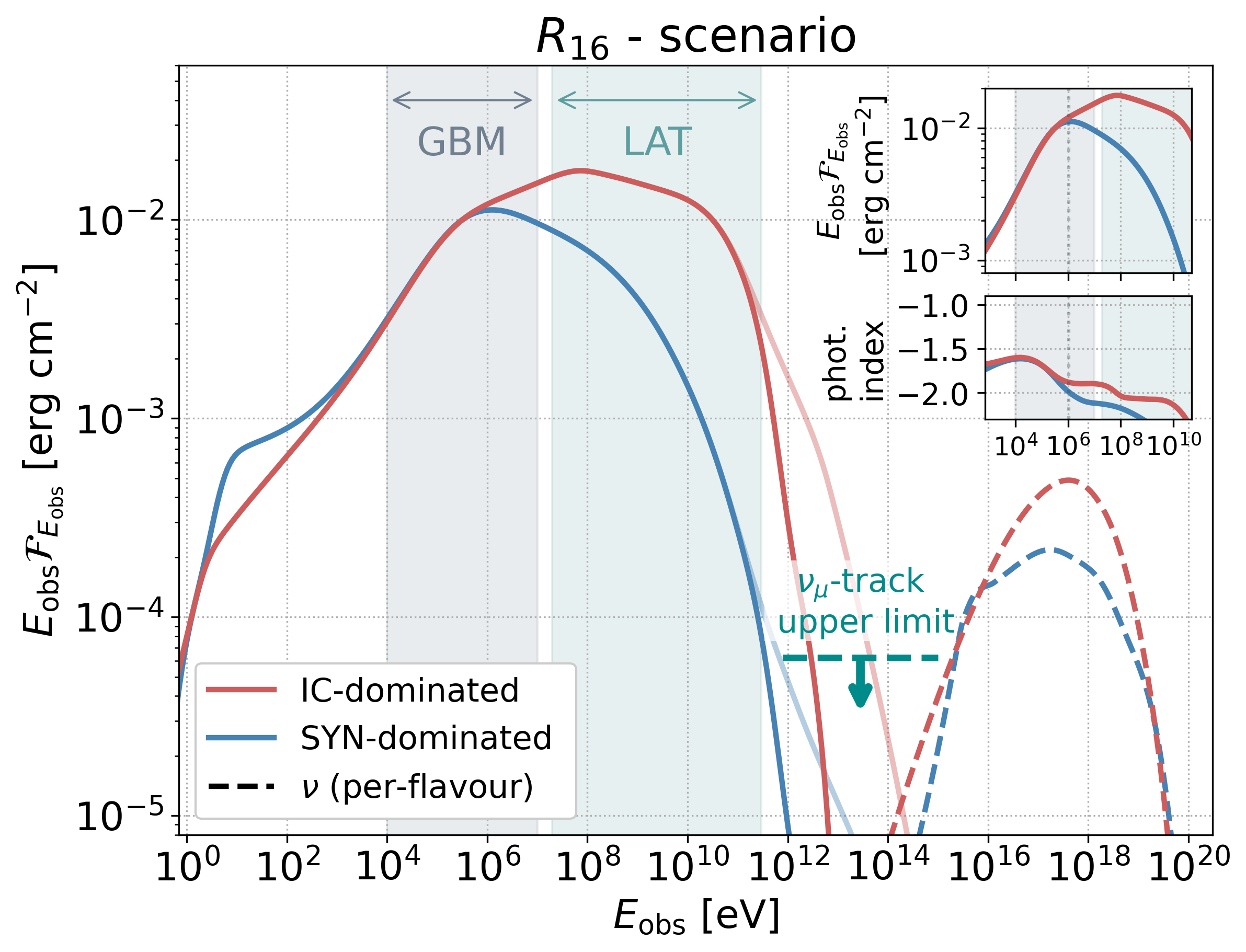}   
\includegraphics[width = 0.48 \textwidth]{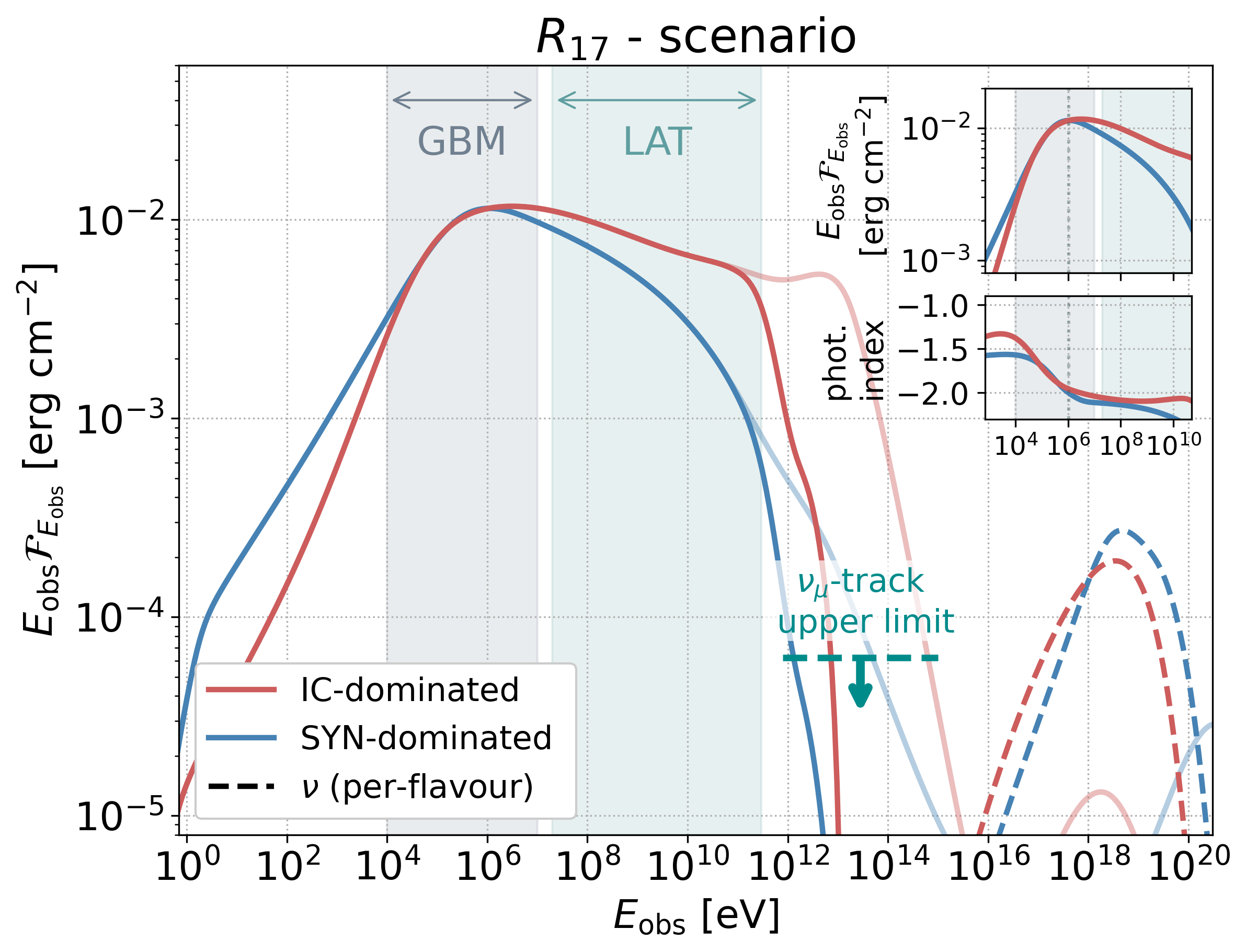}
    \caption{Modelled spectra $E_\mrm{obs} \mathcal{F}_\mrm{E_{obs}}$ for the (left) ``\Rlow'' and ``\Rhigh'' (right) scenarios, in both cases including a SYN- and a IC-dominated scenario (for parameters see \reftab{input_params}). Transparent curves correspond to the fluence without taking EBL attenuation into account, dashed curves to the per-flavour neutrino fluences. The latter are compared with the IceCube upper limit reported in \citet{GRB221009A_IceCube_GCN32665}.
    The inset shows a zoom-in to the spectrum around the peak in the \fermi \, GBM/LAT range (indicated as a shaded region in all plots), with a dotted vertical line at 1060~keV to indicate the position of the observed peak. For this energy range we further indicate the photon indices, that are defined of the slope of d$N_\mrm{ph}$/d$E_\mrm{obs}$. }
    \label{fig:spectra}
\end{figure*}

\textit{Photon spectra.} Fig.~\ref{fig:spectra} shows the predicted time-integrated photon fluences as a function of observed energy; the inset is a zoom-in on the spectra around the peak including the photon index (that is defined as the spectral slope of d$N_\mrm{ph}$/d$E_\mrm{obs}$).

We first discuss the results for the SYN-dominated cases (blue lines in both panels). The spectrum around the peak is by construction independent of the typical collision radius and given by the standard synchrotron fast-cooling predictions: approaching a photon index $-3/2$ below the peak and $-(p_{\rm e}+2)/2$ above the peak. 
Differences between ``\Rhigh'' and ``\Rlow'' are visible at low and high energies with respect to the MeV peak. For ``\Rlow'', where the densities are higher, emission of secondaries increases the fluence in the eV range, leading to a low-energy spectral break. However, the spectrum in the LAT band is still dominated by the synchrotron emission of fast-cooling primaries. For ``\Rhigh'', the power-law spectrum with photon index $-1.5$ extends to the lowest energies (till synchrotron self-absorption becomes important), while higher maximal synchrotron energies yield a harder high-energy spectrum. 

In the ``\Rlow'' IC-dominated case (red curves in left panel), the spectral slope below the peak is still $-1.5$, but the HE emission (in the LAT range) is modified by the ICS emission of primary and secondary leptons that creates an almost flat spectrum (i.e. photon index $\sim -2$). In this case, the peak of the broadband spectrum is shifted to $\sim 1$~GeV.  
In the ``\Rhigh'' IC-dominated case (red curves in right panel) the photon index below the MeV peak (but not within the GBM band) becomes asymptotically $\sim -1.25$, which is the highest value that can be produced in our model. This is indicative of electron cooling via ICS in the Klein-Nishina regime \citep{Daigne:2010fb}.
In the LAT band the spectrum is a power-law that extends to about 10 TeV (where the Klein-Nishina ICS emission of primary electrons with Lorentz factor $\gamma'_{\rm e, min}$ dominates). However, due to EBL attenuation this spectral feature is washed out. We note that in the ``\Rhigh'' IC-dominated case the contribution of secondary leptons in the GBM and LAT bands is negligible.
For a more detailed discussion on spectra and their decomposition, we point to Sec.~4 and 5 in \citet{Rudolph:2022ppp}. 

In addition to the full time-integrated spectra, we computed the spectra for the three main emission periods, namely the first two bright pulses and the late-time pulse (not explicitly shown). Finding little difference in the spectra for these three pulses we predict no significant spectral evolution during these three emission periods.

\textit{Neutrinos.} We include in Fig.~\ref{fig:spectra} the predicted neutrino spectra (per flavor)\footnote{The synchrotron cooling of pions and muons is taken into account, see more detailed discussion in \citet{Rudolph:2022ppp}.} and the corresponding IceCube limits \citep{GRB221009A_IceCube_GCN32665} for the different scenarios as dashed curves. We also compute the number of expected neutrino events in IceCube with the appropriate point-source effective area for the declination range 18-21 degrees \citep{IceCube:2021xar}, and find $n_{\nu_\mu}  = 0.012$ (0.006) for the ``\Rhigh'' SYN- (IC-) dominated case, and  $n_{\nu_\mu}  = 0.17$ (0.29) for the ``\Rlow'' SYN- (IC-) dominated case. The predicted neutrino fluences are thus below the IceCube limits in all cases and consistent with non-detection, as a result of the relatively large $R_{\mathrm{Coll}}$ paired with the chosen baryonic loadings in consistency with the findings in one zone models~\citep{Ai:2022kvd,Murase:2022vqf}. For ``\Rlow'', however, a baryonic loading of $f_\mrm{p/e} \gg 3$ is expected to be in tension with neutrino limits, and if observations eventually favor a variability timescale $\delta t_\mrm{var} \lesssim 1$~s the baryonic loading would be limited to an even smaller value.

Contrary to \citet{Rudolph:2022ppp}, we normalise the initial engine kinetic energy to achieve the same $\gamma$-ray isotropic energy, which for the IC-dominated scenarios increases the required energy (see Tab.~\ref{tab:input_params}). This increases the energy transferred to non-thermal protons, and subsequently the neutrino fluences. The effect can be noticed clearly for the ``\Rlow''-scenario. For the ``\Rhigh''-scenario the neutrino production efficiency is limited by the low(er) maximal proton energies, due to the lower magnetic fields obtained in the IC-dominated case.
It is interesting that the peak neutrino energies of $10^{17}$ to $10^{19} \, \mathrm{eV}$ exceed the expectation of the standard neutrino model for GRBs ($10^{15} \, \mathrm{eV}$, see e.g. \citet{Hummer:2011ms}) by at least two orders of magnitude in energy. This is a result of the synchrotron-cooling dominated spectral indices below the peak (the photon number density peaks at lower energies) paired with weak magnetic field effects on the secondaries as a consequence of large $R_{\mathrm{Coll}}$. Therefore, energetic GRBs may be a target for future radio detection experiments.

\section{Discussion}

There are a number of effects which can be 
included in order to enrich model.
For example, the different pulses may be produced by collisions of shells with very different Lorentz factor ranges or even microphysics parameters. This can cause a combination of SYN- and IC-dominated scenarios in different peaks \citep[see also][for the reverse shock]{Zhang:2022lff}, or suppression of VHE emission in others (by low Lorentz factors enhancing the $\gamma \gamma$ optical thickness, and also the neutrino production). One may speculate that a non-observation of the late-term peak by \fermi-LAT or a contribution to the LHAASO signal could be produced by such effects.
Within each peak, low Lorentz factors and a strong correlation between collision radius and observation time can cause an early suppression VHE photons, which may be interpreted as a delay~\citep{Bustamante:2016wpu}. 
While our spectral index above the peak can be adjusted by the electron injection index and the efficiency of ICS, our model may not accommodate very hard low-energy photon indices of $\sim -1$, unless additional components, such as photospheric thermal emission, are considered. In lepto-hadronic models these components would increase the number of target photons available for photo-hadronic interactions. We note that the highest photon index of $-1.25$ was obtained in the ``\Rhigh'' IC-dominated scenario, but well below the GBM band.
Whether such a high dissipation radius is indeed realistic should be confirmed by observations of the variability timescale and estimates of the Lorentz factor.

The GRB central engine may be a newly formed accreting black hole or magnetar. For energetic bursts, such as GRB~221009A, the latter scenario can be excluded because the available energy is limited by the magnetar’s rotational energy to $\sim 2\cdot 10^{52}$~erg \citep{1992Natur.357..472U, 2004ApJ...611..380T}. The rotational energy, $E_{\rm rot}$, of an accreting black hole can be extracted via electromagnetic fields, provided there is a strong large-scale magnetic field threading the black hole horizon \citep[Blandford-Znajek mechanism (BZ)][]{Blandford:1977ds}. The isotropic equivalent energy of the jet can be written as $E_{\rm jet, iso}=\eta_j f_{\rm b}^{-1} E_{\rm rot}$, where $\eta_j<1$ is the fraction of rotational energy ending up in the jet, $f_{\rm b}=1 - \cos(\theta_j)\approx \theta_j^2/2$ is the beaming factor, $\theta_j$ is the jet half-opening angle, $E_{\rm rot}=f(a) M_{\rm BH} c^2$, $f(a_*)=1-\sqrt{(1+ \sqrt{1-a_*^2})/2}$, and $a_*$ is the dimensionless black hole spin. Adopting $\theta_j = 3.5$~deg \citep{2022GCN.32755....1D}\footnote{This was obtained using a prompt-phase efficiency of 0.2. In our model this efficiency is lower, which would require either a larger density of the surrounding medium or yield a smaller opening angle.} and $\eta_j=0.5$ we find that a maximally spinning ($a_*=1$) black hole with $M_{\rm BH}=10~M_{\odot}$ can produce $E_{\rm jet,  iso}\simeq 1.4\cdot 10^{57}$~erg, which is comfortably larger than $E_{\rm kin,ini}$ required for the ``\Rlow'' scenario. Moderate spins ($a_*\sim 0.5$) would require $M_{\rm BH} \sim 40~M_{\odot}$ to meet the model's energetic requirements. The formation of such massive black holes in the collapsar scenario for long-duration GRBs (with stellar masses $<40~M_\odot$ in the zero-age main sequence) is not expected \citep{1993ApJ...405..273W} -- see, however, \cite{2021arXiv211103094S} for very massive collapsars. The ``\Rhigh''-scenario with baryonic loading 30 corresponds to $E_{\rm kin,ini}\ge (3-5)\cdot 10^{57}$~erg which would need $a_*=1$ and $M_{\rm BH}\ge 20~M_\odot$. Hence, this scenario is unlikely on energetic grounds.
We point out that although a BZ-powered jet would be initially Poynting-flux dominated, our internal shock model implies a matter-dominated jet at the dissipation region

\citep[see e.g.][for possible scenarios for energy conversion in jets]{Granot:2010en, Giannios:2018nin, Gottlieb:2022tkb}. For a non-negligible magnetisation at the dissipation radius, a different mechanism for particle energization may instead be invoked, such as magnetic reconnection \citep[\textit{e.g.}][]{Sironi:2014jfa, Guo:2015ydj, Werner:2016fxe}.

The energy budget is further typically constrained from the afterglow brightness. Here it is noteworthy to mention that a part of the energy in the afterglow being dissipated in VHE may relax the requirements on the prompt-phase efficiency, especially for proton-synchrotron models for the VHE emission \citep{Isravel:2022glo}. 
Note that a part of the afterglow energy going into thermal particles may equally increase the allowed kinetic energy of the blastwave after the prompt phase. 
Afterglow observations are also used to infer typical parameters of the GRB, for example \citet{Ren:2022icq} find a Lorentz factor $\Gamma_0 = 190$ (with room for slightly higher $\Gamma_0$, see their Fig.~3) for the afterglow. This would favour our ``\Rlow''-scenario, that has a typical Lorentz factor $\langle \Gamma_\mrm{fin} \rangle \sim 230 $ after the prompt phase. 

Our obtained (averaged) maximal proton energies are in the range between $10^{20}$ and $2 \cdot 10^{21} \, \mathrm{eV}$ under the assumption of efficient particle acceleration, see Table~\ref{tab:input_params}. Higher proton energies are expected in the SYN-dominated scenarios than the IC-dominated ones because the acceleration rate is higher due to higher magnetic fields. Our maximal energies are compatible with the values used e.g. in \citet{Das:2022gon} (100 EeV) to describe the LHAASO VHE photons from EBL interactions. We note that the time delay induced by the extragalactic magnetic fields (EGMFs) requires extremely low field values paired with large proton energies, which means that this challenge can be somewhat mitigated in our SYN-dominated scenario by the high proton energies. We also note that the EGMF induced delay is very large (assumed to be limited to the observed LHAASO window of 2000s) even under aggressive assumptions for the magnetic field, which means that the protons must be accelerated in the prompt phase of the GRB and further delays induced by the afterglow cannot be accommodated. We find that most of the UHECR protons are emitted within the first 100s in our model, which is compatible with this picture. Our obtained maximal proton energies are also higher than the maximum corresponding rigidity $R_{\mathrm{max}} \simeq 1-3 \, \mathrm{EV}$ required to describe UHECR data \citep{Heinze:2019jou} -- where details depend on the assumed cutoff shape.

\section{Summary and conclusions}

In this letter we have presented a state-of-the-art multi-messenger emission model for the prompt emission of GRB 221009A. In this model, plasma shells are ejected from a central engine with varying Lorentz factors; these eventually catch up and energy is dissipated in internal shocks. Our radiation model includes the effect of UHECR protons self-consistently, such as the electromagnetic cascade in each shell, generated from secondary electrons, positrons and photons produced in photo-hadronic interactions. Our assumptions for the baryonic loading (3 and 30) have been motivated by the paradigm that energetic GRBs, such as GRB 221009A, could be sources of the UHECRs; they are also consistent with the hypothesis that the highest-energy LHAASO photons come from interactions of UHECRs with the EBL. 

We have demonstrated that an intermittent engine can reproduce the observed prompt light curves if a quiescent period of about 200 seconds is included and assuming a variability timescale of $\sim 1$s. We have implemented relatively large Lorentz factors and therefore collision radii supported by various arguments (such as $\gamma \gamma$ optical thickness, neutrino non-observation). 
Our predicted electromagnetic spectra exhibit synchrotron fast cooling dominated spectral indices below the MeV peak, except for an ICS dominated scenario with energy dissipation at high radii that yielded a softer spectral index. Above the peak, ICS can affect the spectral index very strongly, an effect which can extend up to the highest energies and enhance the VHE fluxes. On the other hand, hadronic contributions did not significantly alter the photon spectra. Since pile-up effects affect the analysis for nearly all instruments due to the high brightness of the GRB, the predictive power of our model may be useful. The predicted neutrino emission was consistent with the non-observation of neutrinos by IceCube due to high predicted peak energies in the range interesting for radio neutrino telescopes. For lower emission radii (due to variability on shorter timescales and/or lower Lorentz factors) the baryonic loading would likely be constrained to lower values than the ones assumed here.
Our findings are consistent with the available rotational energy which can be extracted from a maximally spinning black hole with a mass of the order of $10 \, M_\odot$ if the baryonic loading is not too far away from energy equipartition; our standard assumption of a baryonic loading of 3 is consistent with this picture.

We conclude that GRB 221009A is an interesting object to test the internal shock model and the paradigm that energetic GRBs could be the sources of UHECRs. While direct signatures of cosmic rays (such as neutrinos) have not been seen, the LHAASO observation of TeV photons could point towards UHE proton acceleration. Our model connects the different messengers for the prompt phase of this GRB.

\textit{Note. --}  During completion of this work \citet{Liu:2022mqe} appeared. In contrast to their paper, we self-consistently model the photon spectra and account for several emission regions along the jet. Their low(er) emission radii are the consequence of a variability timescale of $82 \, \mathrm{ms}$ that was derived from an analysis of the GBM light curve up to $219 \, \mathrm{s}$, whereas we inferred the variability timescale from the \spiacs \, light curve also including the bright emission period most relevant for the spectra. Overall, their limits on the baryonic loading are compatible with our findings. 

\acknowledgments
We would like to thank the anonymous referee for a constructive report and their intuitive comments.
We would like to thank Irene Tamborra, Iftach Sadeh and Marc Klinger
for reading the manuscript and useful comments and Sylvia Zhu and Andrew Taylor for discussions around GRB~221009A.
A.R. received funding from the Carlsberg Foundation (CF18-0183). M.P. acknowledges support from the MERAC Fondation through the project THRILL and from the Hellenic Foundation for Research and Innovation (H.F.R.I.) under the ``2nd call for H.F.R.I. Research Projects to support Faculty members and Researchers'' through the project UNTRAPHOB (Project ID 3013).
 
\vspace{5mm}

\software{\soft{gammapy} \citep{Deil:2017yey, Nigro:2019hqf} and Python v3.9.}

\bibliography{references}

\end{document}